\documentclass{article}

\usepackage[dblblindworkshop,final]{neurips_2026}
\workshoptitle{Real-Time Multimodal Conversational AI}

\usepackage{amsmath}
\usepackage[utf8]{inputenc} 
\usepackage[T1]{fontenc}    
\usepackage{hyperref}       
\usepackage{url}            
\usepackage{booktabs}       
\usepackage{amsfonts}       
\usepackage{nicefrac}       
\usepackage{microtype}      
\usepackage{xcolor}         
\usepackage{graphicx}

\title{From Spectrograms to Shallow Speech SSL:
Audio-Frontend Design for Generator-Disjoint and Low-Latency Audio--Visual Deepfake Detection}

\author{%
  Qiwu Wen \\
  Luqia Technologies \\
  \texttt{qiwu.wen@luqia.ca}
  \And
  Jahangir Alam \\
  Luqia Technologies \\
  \texttt{jahangir.alam@luqia.ca}
  \And
  Eric Charton \\
  Luqia Technologies \\
  \texttt{eric.charton@luqia.ca}
}

\begin{document}
\maketitle
\begin{abstract}
Video-mediated conversational systems rely on synchronized speech and facial
video, making audio--visual input integrity important for reliable perception.
We study audio-frontend design in a fixed MRDF
(Modality-Regularization-based DeepFake)-based pipeline with AV-HuBERT visual
processing and an MSCT (multi-scale cross-modal transformer) fusion encoder.
We compare spectrogram-style log-filterbank features with frozen wav2vec~2.0
and WavLM representations, dual-SSL fusion, executed depth, and shallow-layer
aggregation. Across three training runs, full wav2vec~2.0 improves mean macro
AUROC from $84.19\%$ to $87.19\%$ on generator-disjoint MMDF.
Six-layer execution further improves detection, and learned non-uniform
aggregation adds $2.53$ percentage points, reaching $90.61\%$.
Shallow execution reduces measured inference latency relative to full-depth
wav2vec~2.0 across the tested input durations; first-second evaluation reaches
$86.85\%$ macro AUROC.
In a single-run comparison, dual-SSL concatenation outperforms either single
stream, while one-way cross-attention yields lower detection performance and
higher latency than concatenation.
Layer-weight analysis shows greater emphasis on early speech layers, while
removing the selector's visual input has little effect on detection performance.
\end{abstract}

\section{Introduction}
\label{sec:introduction}

Video-mediated conversational systems perceive users through speech, facial
motion, remote video, and increasingly realistic virtual avatars.
Manipulated talking-head video or inconsistent audio--visual behavior can
therefore corrupt perceptual input before downstream dialogue reasoning or
response generation begins.
Audio--visual (AV) deepfake detection provides a natural integrity check for
this setting, but practical perception requires both generalization to new
synthesis methods and computational feasibility under a low-latency budget.

Recent AV detectors strengthen cross-modal correspondence, representation
regularization, or multimodal fusion.
Cross-Modality and Within-Modality Regularization for Audio-Visual DeepFake
Detection (MRDF) preserves modality-specific and multimodal evidence
~\cite{Zou2024CrossModality};
Audio-Visual Feature Fusion (AVFF) learns intrinsic AV correspondence
~\cite{Oorloff2024AVFF};
and the multi-scale cross-modal transformer encoder (MSCT) introduces
multi-scale self-attention (MSSA) and differential cross-modal attention
(DCA)~\cite{Wei2026MSCT}.
SpeechForensics further shows that speech representations can expose
inconsistencies between spoken content and facial behavior
~\cite{Liang2024SpeechForensics}.
These advances improve how audio and video interact, but leave a more basic
choice comparatively underexamined:
\emph{what audio representation should a strong AV detector receive?}

To isolate this question, we retain the AV-HuBERT visual frontend used by the
official MRDF code~\cite{Shi2022AVHuBERT}, its classification and alignment
objectives, and our implementation of the MSCT MSSA--DCA fusion modules
(Appendix~\ref{frameworkDetails}).
The inherited MRDF audio path constructs spectrogram-style tokens from
stacked log-filterbank features.
Self-supervised learning (SSL) provides alternatives directly from waveform:
wav2vec~2.0 (W2V2) learns contextual speech representations
~\cite{Baevski2020Wav2vec2}, while WavLM targets a broader range of acoustic
and speaker-related tasks~\cite{Chen2022WavLM}.
For audio-only deepfake detection, \emph{Two Views, One Truth} shows that
heterogeneous spectral and SSL views can be fused through concatenation or
cross-attention~\cite{ElKheir2025TwoViews}; layer-wise analysis further
reports that lower SSL layers can preserve strong forensic cues with less
computation~\cite{ElKheir2025Layerwise}.
A recent AV representation study likewise finds useful pretrained features
but reports remaining generalization limitations~\cite{Boldisor2026SSLAV}.
We therefore keep the AV detector architecture unchanged and study frontend
replacement, fusion between frozen speech views, executed SSL depth, and
shallow-layer aggregation.

MMDF provides a modern generator-disjoint setting spanning GAN-,
diffusion-, Diffusion-Transformer-, and flow-matching-based synthesis
pipelines, with disjoint seen and unseen generators
~\cite{Kim2026XAVDT}.
Training and validation use FaceAdapter, Hallo2, and LivePortrait, while final evaluation
uses AniPortrait, Hunyuan, and MegActor-S.
In our protocol, the audio auxiliary target is bona fide for both real and
manipulated clips, so speech acts primarily as a reference for visual and
cross-modal forgery reasoning.
We ask:
\textbf{RQ1} whether frozen speech SSL and multi-view fusion improve over the
inherited spectrogram frontend;
\textbf{RQ2} how executed SSL depth affects detection and latency; and
\textbf{RQ3} whether learned non-uniform aggregation of shallow layers
improves detection at nearly fixed runtime.
Our results support shallow SSL execution and learned layer aggregation,
while showing that combining SSL streams depends on the fusion rule.
Appendix~\ref{additional_part} provides additional statistical, layer-weight,
latency, and prefix analyses.

\section{Audio-Frontend Design}
\label{sec:frontends}

\begin{figure}[t]
\centering
\includegraphics[width=\linewidth]{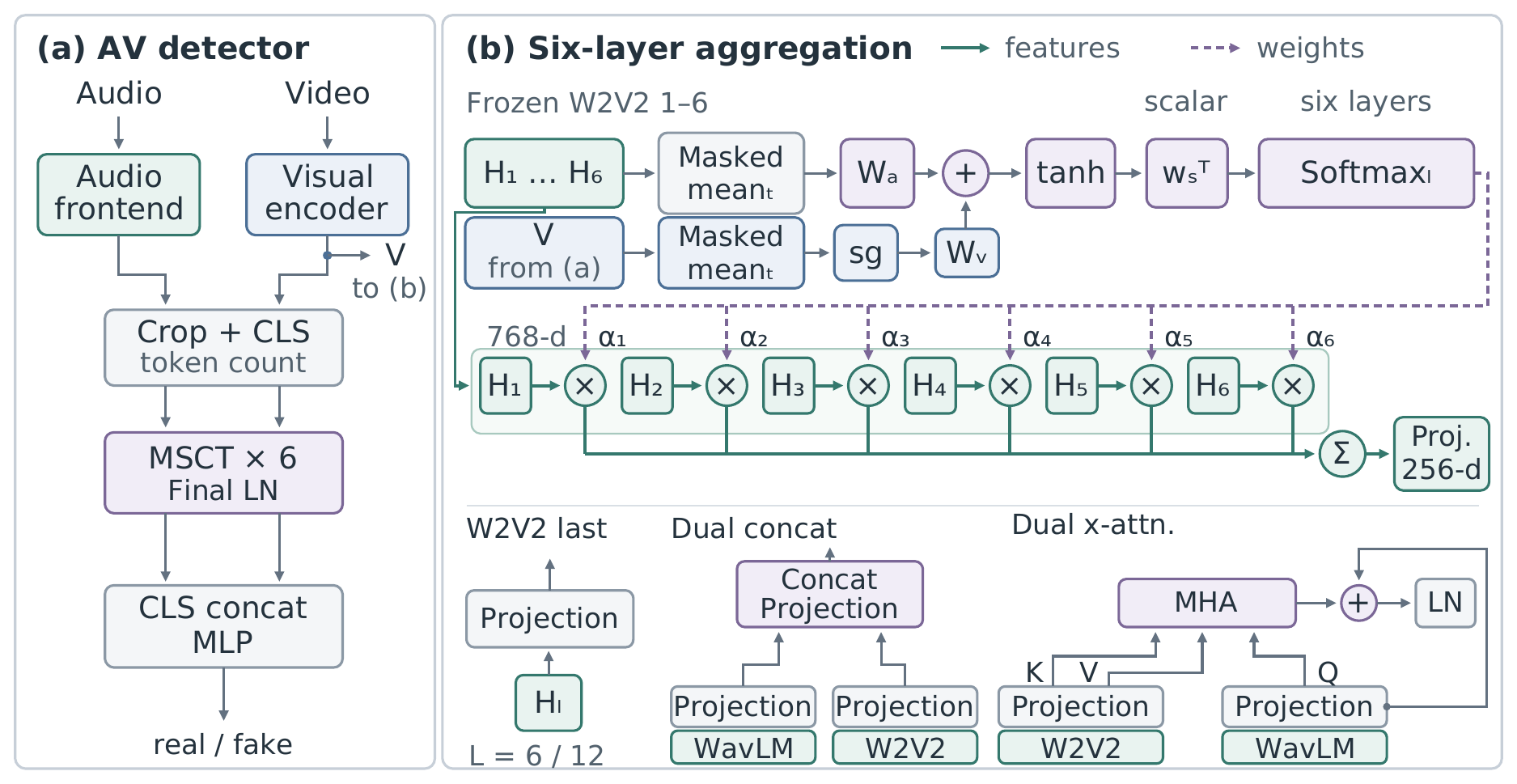}
\caption{(a) Fixed AV detector. (b) W2V2-L6-weighted and comparison
frontends. Audio and stop-gradient visual summaries produce weights for
the original layer sequences. Q/K/V label the MHA inputs.
Solid arrows carry features; dashed arrows carry weights.
Alignment, frontend input/output normalization, biases, dropout, masks,
and auxiliary heads are omitted.
Dual-SSL comparisons appear in Appendix~\ref{dual_ssl}.}
\label{fig:method}
\end{figure}

\paragraph{Fixed AV backbone.}
All configurations use the same MRDF-based training and regularization
pipeline~\cite{Zou2024CrossModality}, with AV-HuBERT visual
processing~\cite{Shi2022AVHuBERT} and a six-block MSCT-style MSSA--DCA
fusion encoder~\cite{Wei2026MSCT}.
Only the audio frontend and its integration into the MSCT token space vary
across experiments; the visual-processing architecture, fusion backbone,
training objectives, and classification-head design are shared.
Figure~\ref{fig:method} shows the detector and shallow aggregation;
Appendix~\ref{frameworkDetails} explains the MSCT implementation with
Figure~\ref{fig:msct}.

\paragraph{Executed depth and shallow-layer aggregation.}
For waveform $x_i^a$, let
$\{\mathbf H^a_{i,1},\ldots,\mathbf H^a_{i,L}\}
=E_{\mathrm{SSL}}^{1:L}(x_i^a)$
denote the hidden sequences produced by the first $L$ Transformer blocks.
Execution stops after block $L$; upper blocks remain loaded but are not
executed. We compare $L\in\{4,6,9,12\}$ and use the six-layer configuration
to compare weighting rules at a fixed half-depth budget.
At a fixed depth,
\begin{equation}
 \widetilde{\mathbf H}^{a}_{i}
 =
 \sum_{\ell=1}^{L}\alpha_{i,\ell}\mathbf H^{a}_{i,\ell},
 \qquad
 \mathbf A_i =
 \operatorname{Norm}\!\left[
   \operatorname{Align}\!\left(P_a\widetilde{\mathbf H}^{a}_{i}\right)
 \right],
 \label{eq:aggregation}
\end{equation}
where $P_a$ projects to the MSCT dimension.
We compare final-layer, uniform, global, audio-conditioned, and
audio--visual-conditioned weights.
\emph{W2V2-L6-weighted} uses audio--visual-conditioned weights over the first
six layers. Appendix~\ref{weights} gives the scorer and examines the learned
weights and the effect of visual conditioning.

\paragraph{Dual-SSL fusion and training.}
We also combine frozen full-depth WavLM and W2V2 through channel
concatenation or one-way cross-attention
(WavLM queries; W2V2 keys and values).
The two configurations differ only in the fusion operator; exact equations
and results are in Appendix~\ref{dual_ssl}.
All models retain the same audio, visual, and multimodal cross-entropy losses
and the MRDF alignment loss, each with unit weight.

\section{Experiments}
\label{sec:experiments}

\paragraph{Protocol and metrics.}
All configurations share the same MMDF splits, video processing, detector,
losses, and optimization settings.
We train for at most 30 epochs with Adam, learning rate $10^{-4}$, weight
decay $10^{-4}$, FP16 automatic mixed precision, and batch size 60.
Checkpoints are selected using generator-macro AUROC on the seen-validation
set. Table~\ref{tab:detection} reports means and standard deviations over
three training runs. Additional frontend and component comparisons use one
matched run per configuration unless stated otherwise
(Table~\ref{tab:frontend_overview}, Appendix~\ref{app:frontend_overview}).
Metrics are computed per unseen generator and macro-averaged.
mEER is the mean of generator-specific EERs; Worst is the minimum generator
AUROC within each training run, averaged across runs for the main results.
Appendix~\ref{app:bootstrap} details the paired bootstrap analysis of saved
predictions.

\paragraph{Latency and prefix evaluation.}
Latency profiling uses one checkpoint per configuration on an NVIDIA A40,
batch size one, and FP16 autocast.
Synchronized $1$, $2$, $4$, $8$, and $19.095$\,s inputs are measured in
three independent processes, each with 50 warm-up and 200 timed iterations.
These controlled durations probe computational scaling and are distinct from
the native MMDF test clips, whose synchronized AV content is approximately
two seconds.
Timing runs from predecoded host tensors to logits and excludes media
decoding and resampling.
Appendix~\ref{latency} reports the complete protocol and path audit.
We also evaluate the saved checkpoints on the first synchronized second of
each native test clip (Appendix~\ref{prefix}).

\begin{table}[t]
\centering
\caption{
Generator-disjoint MMDF results (\%).
Values are mean$\pm$standard deviation over three training runs.
Gap is validation mAUC minus test mAUC.
}
\label{tab:detection}
\small
\setlength{\tabcolsep}{4.4pt}
\renewcommand{\arraystretch}{1.03}
\begin{tabular}{lcccc}
\toprule
Frontend
& Test mAUC $\uparrow$
& Gap $\downarrow$
& mEER $\downarrow$
& Worst $\uparrow$\\
\midrule
Spectrogram
& $84.19{\pm}3.40$
& 13.93
& $23.07{\pm}3.40$
& $\mathbf{79.68{\pm}4.68}$\\
W2V2-L12
& $87.19{\pm}3.84$
& 10.75
& $19.02{\pm}3.74$
& $76.12{\pm}6.96$\\
W2V2-L6-last
& $88.08{\pm}3.82$
& 9.45
& $18.00{\pm}3.54$
& $74.28{\pm}9.18$\\
W2V2-L6-weighted
& $\mathbf{90.61{\pm}1.81}$
& $\mathbf{7.20}$
& $\mathbf{15.17{\pm}1.57}$
& $78.58{\pm}5.42$\\
\bottomrule
\end{tabular}
\end{table}

\begin{table}[t]
\centering
\caption{
Duration-dependent p50 model-path latency (ms) on one NVIDIA A40.
Full denotes a controlled 19.095-s synchronized input; slope is a descriptive
fit over all five durations.
}
\label{tab:latency_duration}
\small
\setlength{\tabcolsep}{5.0pt}
\renewcommand{\arraystretch}{1.03}
\begin{tabular}{lrrrr}
\toprule
Frontend
& 1 s
& 8 s
& Full
& Slope ms/s $\downarrow$\\
\midrule
Spectrogram
& $\mathbf{23.15}$ & 33.20 & 52.05 & 1.59\\
W2V2-L12
& 32.28 & 35.80 & 52.93 & 1.14\\
W2V2-L6-last
& 28.61 & $\mathbf{31.70}$ & $\mathbf{46.70}$ & 0.99\\
W2V2-L6-weighted
& 29.90 & 33.24 & 47.37 & $\mathbf{0.96}$\\
\bottomrule
\end{tabular}
\end{table}

\paragraph{Frontend replacement and execution depth.}
Table~\ref{tab:detection} shows that full W2V2 raises mean test mAUC from
$84.19\%$ to $87.19\%$, and six-layer execution further increases it to
$88.08\%$.
L6-last is faster than L12 at every profiled duration; at $19.095$\,s,
p50 decreases from $52.93$ to $46.70$\,ms
(Table~\ref{tab:latency_duration}).
The paired mean detection gain over L12 is $.89$ percentage points, with a
95\% hierarchical bootstrap interval of $[+.38,+1.43]$ points
(Appendix~\ref{app:bootstrap}).

\paragraph{Learned shallow-layer aggregation.}
Weighted L6 reaches $90.61\pm1.81\%$ mAUC, improving on L6-last and
Spectrogram by $2.53$ and $6.41$ percentage points, respectively.
Both gains occur in all three training runs; their 95\% hierarchical
bootstrap intervals are $[+.48,+4.65]$ and $[+4.86,+8.67]$ points
(Appendix~\ref{app:bootstrap}).
For the checkpoint used in the intervention analysis, zeroing or shuffling
the selector's visual summary changes mAUC by less than $.02$ points,
whereas uniform layer weighting reduces it by $1.34$ points
(Appendix~\ref{weights}).
These results associate the gain primarily with learned unequal layer
weights. Appendix~\ref{overlap} examines which errors the weighted model
corrects relative to L6-last.

\paragraph{Input-duration and generator effects.}
The spectrogram path is faster for short inputs, whereas truncated W2V2 has
lower latency at the longest profiled duration.
These measurements include CPU log-filterbank extraction for Spectrogram and
GPU speech encoding for W2V2 (Appendix~\ref{latency}).
Detection gains also differ by generator: weighted L6 improves AniPortrait
and MegActor-S on average, while Hunyuan remains slightly below Spectrogram
(Appendix~\ref{broader}).

\section{Conclusion and Limitations}
\label{sec:conclusion}

In the MRDF-based detector with MSCT fusion, speech-frontend design affects
both unseen-generator detection and inference latency.
Full W2V2 improves average detection over Spectrogram; shallow execution
further improves detection while reducing runtime, and learned layer
weighting adds a gain with little extra computation.
The dual-SSL comparison also favors concatenation over the evaluated
one-way cross-attention configuration.

The evaluation covers one detector family and one generator-disjoint
benchmark, with three-run main results and single-run component comparisons.
Latency and prefix tests start from predecoded inputs and do not evaluate
causal streaming or full conversational decision time.
Unused upper W2V2 blocks remain resident in memory, so reduced execution
depth does not remove their memory cost.

\begin{ack}
Authors wish to acknowledge Canada Economic Development for Quebec Regions
(CED) and the Ministry of Economy, Innovation and Energy (MEIE) of the
Government of Quebec for their continued support.
\end{ack}

\begingroup
\small
\bibliographystyle{plainnat}
\bibliography{refs}
\endgroup


\appendix
\section{Additional Method and Experimental Details}
\label{additional_part}

Single-run comparisons and inference analyses use matched checkpoints from
one training run; tables combining three runs are identified in their
captions. For reproducibility, runs 1--3 correspond to random seeds 12, 22,
and 42, respectively, and the single-run analyses use run 3.

\subsection{Additional Frontend-Design Results}
\label{app:frontend_overview}

\begin{table}[t]
\centering
\caption{
Single-run audio-frontend comparison under the shared MSCT/MRDF detector.
Native denotes the available clip (median synchronized AV duration:
$2.015$\,s); 1-s results use the same Spectrogram/W2V2 checkpoints.
}
\label{tab:frontend_overview}
\small
\setlength{\tabcolsep}{4.2pt}
\renewcommand{\arraystretch}{1.02}
\begin{tabular}{lrrrr}
\toprule
Frontend / fusion
& 1-s mAUC $\uparrow$
& Native mAUC $\uparrow$
& Native mEER $\downarrow$
& Native Worst $\uparrow$\\
\midrule
Spectrogram
& 80.65 & 84.30 & 22.75 & \textbf{80.12}\\
W2V2-L12
& 83.35 & 84.18 & 21.72 & 69.08\\
WavLM-L12
& -- & 80.11 & 27.77 & 70.56\\
Dual SSL concat
& -- & 89.00 & 17.46 & 76.97\\
Dual SSL cross-attn.
& -- & 84.72 & 21.87 & 69.09\\
\midrule
W2V2-L6-last
& 82.67 & 84.80 & 21.03 & 66.25\\
W2V2-L6-weighted
& \textbf{86.85} & \textbf{89.60} & \textbf{16.08} & 75.36\\
\bottomrule
\end{tabular}
\end{table}

\paragraph{Frontend comparison.}
Table~\ref{tab:frontend_overview} compares the audio representations.
Weighted shallow W2V2 and dual-SSL concatenation reach $89.60\%$ and
$89.00\%$ native mAUC, compared with $84.30\%$ for Spectrogram.
Full W2V2 is close to Spectrogram in this run, while WavLM scores lower.
Replacing concatenation with the evaluated cross-attention reduces dual-SSL
mAUC to $84.72\%$.
The fusion rules obtain similar seen-validation scores ($97.92\%$ versus
$97.40\%$), but differ more on unseen generators
(Appendix~\ref{dual_ssl}).
Weighted shallow W2V2 reaches $86.85\%$ from the first second, $2.75$ points
below its native-clip score; Appendix~\ref{prefix} gives the generator-wise
interpretation.
Tables~\ref{tab:detection} and~\ref{tab:latency_duration} report the
three-run detection comparison and duration-dependent latency.

\subsection{Paired Bootstrap and Training-Run Consistency}
\label{app:bootstrap}
We use paired bootstrap resampling of saved test predictions, with the same
resampled examples for both models and stratification by generator and class.
Run-specific intervals condition on a fixed checkpoint; hierarchical
intervals additionally resample the three training runs.
These hierarchical intervals remain exploratory because training variability
is estimated from three runs.

\begin{table}[h]
\centering
\caption{
Paired mAUC gains in percentage points across three training runs.
Run-wise gains follow the run order defined at the start of the appendix.
}
\label{tab:bootstrap}
\small
\setlength{\tabcolsep}{4.0pt}
\begin{tabular}{lcccc}
\toprule
Comparison
& Run-wise gains
& Mean
& Positive runs
& 95\% CI\\
\midrule
L12 $-$ Spectrogram
& $+5.13,+3.97,-.12$
& $+2.99$ & $2/3$ & $[+.06,+5.14]$\\
L6-last $-$ L12
& $+1.29,+.76,+.63$
& $+.89$ & $3/3$ & $[+.38,+1.43]$\\
L6-weighted $-$ L6-last
& $+2.37,+.43,+4.79$
& $+2.53$ & $3/3$ & $[+.48,+4.65]$\\
L6-weighted $-$ Spectrogram
& $+8.78,+5.16,+5.30$
& $+6.41$ & $3/3$ & $[+4.86,+8.67]$\\
\bottomrule
\end{tabular}
\end{table}

Table~\ref{tab:bootstrap} shows positive gains in all three runs for both
shallow execution and learned aggregation, compared with two runs for full
W2V2 replacement.
The weighting gain over L6-last ranges from $.43$ to $4.79$ points; the
sample-level interval for the run with the smallest gain includes zero.

\subsection{Weighted Aggregation and Conditioning Interventions}
\label{weights}

W2V2-L6-weighted computes one scalar weight per executed speech layer from
pooled audio and visual features:
\begin{equation}
\begin{aligned}
 \bar{\mathbf h}^{a}_{i,\ell}
 &=\operatorname{MaskedMean}(\mathbf H^{a}_{i,\ell}),\qquad
 \bar{\mathbf h}^{v}_{i}
 =\operatorname{MaskedMean}(\mathbf V_i),\\
 \mathbf r_{i,\ell}
 &=\tanh\!\left(
   W_a\bar{\mathbf h}^{a}_{i,\ell}
   +W_v\operatorname{sg}(\bar{\mathbf h}^{v}_{i})
   +\mathbf b
 \right),\\
 e_{i,\ell}
 &=\mathbf w_s^\top\mathbf r_{i,\ell},\qquad
 \alpha_{i,\ell}
 =\operatorname{softmax}_{\ell}(e_{i,\ell}),
\end{aligned}
\label{eq:appendix_weighting}
\end{equation}
where $\operatorname{sg}$ denotes stop-gradient.
Aggregation is followed by a $768\!\rightarrow\!256$ projection,
length-aware temporal alignment, and two 256-dimensional LayerNorms.

In Figure~\ref{fig:method}b, $\mathbf V_i$ denotes the projected visual
tokens before the detector's common-token-count crop. The scorer is shared
across layers, and each sample's six scalar weights are broadcast over the
time and channel dimensions of the original hidden sequences.

\begin{table}[h]
\centering
\caption{
Mean unseen-test weights and effective number of active layers for three
training runs.
}
\label{tab:weights}
\small
\setlength{\tabcolsep}{3.7pt}
\begin{tabular}{lcccccccc}
\toprule
Run
& $\alpha_1$ & $\alpha_2$ & $\alpha_3$
& $\alpha_4$ & $\alpha_5$ & $\alpha_6$
& $H_{\mathrm{norm}}$ & $N_{\mathrm{eff}}$\\
\midrule
1 & .269 & .216 & .181 & .120 & .109 & .106 & .958 & 5.57\\
2 & .386 & .231 & .146 & .095 & .076 & .066 & .878 & 4.84\\
3 & .430 & .208 & .151 & .096 & .061 & .054 & .846 & 4.57\\
\midrule
Mean & .362 & .218 & .159 & .103 & .082 & .076 & .894 & 4.99\\
\bottomrule
\end{tabular}
\end{table}

Table~\ref{tab:weights} shows greater weight on earlier layers while all
six layers contribute. Across runs, the mean weight is $.362$ for the first
layer and $.076$ for the sixth.
The first layer receives the largest weight for $88.0\%$, $98.1\%$, and
$99.5\%$ of test samples in the three runs, respectively.

\begin{table}[h]
\centering
\caption{
Weighting-rule interventions on one fixed checkpoint.
The shuffled result averages five fixed within-group permutations.
}
\label{tab:interventions}
\small
\setlength{\tabcolsep}{7.0pt}
\begin{tabular}{lrrr}
\toprule
Inference rule
& mAUC
& $\Delta$ mAUC
& Worst AUC\\
\midrule
Normal & 89.598 & --- & 75.362\\
Zero visual contribution & 89.613 & $+.015$ & 75.416\\
Shuffled visual summary (mean) & 89.615 & $+.017$ & 75.389\\
Fixed validation-mean weights & 89.418 & $-.180$ & 74.949\\
Uniform weights & 88.259 & $-1.338$ & 72.087\\
\bottomrule
\end{tabular}
\end{table}

Table~\ref{tab:interventions} compares changes to the weighting rule while
keeping the main visual detection branch active.
Removing or shuffling the selector's visual summary changes mAUC by less
than $.02$ points. Fixed validation-mean weights reduce mAUC by $.18$
points, whereas uniform weights reduce it by $1.34$ points, including a
$3.27$-point decrease on Hunyuan.
For this checkpoint, a learned unequal mixture of speech layers preserves
most of the gain even when the weights are fixed across samples.

\subsection{Duration-Dependent Inference Latency}
\label{latency}

\begin{table}[h]
\centering
\caption{
Duration-dependent p50 latency (ms).
}
\label{tab:latency_full}
\scriptsize
\setlength{\tabcolsep}{3.1pt}
\begin{tabular}{lrrrrrrrr}
\toprule
Frontend
& 1 s & 2 s & 4 s & 8 s & 19.095 s
& Slope & Intercept & $R^2$\\
\midrule
Spectrogram
& 23.15 & 24.92 & 27.93 & 33.20 & 52.05
& 1.59 & 21.43 & .998\\
W2V2-L12
& 32.28 & 33.29 & 34.29 & 35.80 & 52.93
& 1.14 & 29.95 & .950\\
W2V2-L6-last
& 28.61 & 29.45 & 30.67 & 31.70 & 46.70
& .99 & 26.64 & .951\\
W2V2-L6-weighted
& 29.90 & 30.95 & 32.01 & 33.24 & 47.37
& .96 & 28.18 & .957\\
\bottomrule
\end{tabular}
\end{table}

Table~\ref{tab:latency_full} reports the complete duration-dependent p50
profile.
At 19.095\,s, p95 latency is $55.77$, $54.57$, $48.12$, and
$48.85$\,ms, respectively.
Peak allocated memory is approximately $556$\,MB for Spectrogram and
$938$--$939$\,MB for W2V2 variants.

The measured spectrogram path includes CPU
\texttt{python\_speech\_features.logfbank}, NumPy conversion, tensor
construction, and a batch-size-one Python loop. W2V2 uses FP16 GPU inference
without resampling. These timings therefore compare the implemented
processing paths under the same measurement boundary.
The controlled $19.095$-s input measures latency scaling, while the native
clips used for detection contain approximately $2$\,s of synchronized AV
content (Appendix~\ref{prefix}).

\subsection{MRDF--MSCT Detection Framework}
\label{frameworkDetails}

\paragraph{MRDF--MSCT pipeline.}
All experiments use the same MRDF-based audio--visual detection
pipeline~\cite{Zou2024CrossModality}.
The visual stream follows the AV-HuBERT-based frontend used in the MRDF
implementation~\cite{Shi2022AVHuBERT}, while the original MRDF fusion
transformer is replaced by six MSCT-style fusion blocks
~\cite{Wei2026MSCT}.
MRDF supplies modality-specific audio and visual classification objectives,
a multimodal classification objective, and cross-modal alignment
regularization.
In this work, the audio frontend is the experimental variable: spectrogram
or SSL representations are projected and temporally aligned into the same
256-dimensional MSCT token space, while the remaining detector structure is
kept fixed.
We refer to this MRDF-based detector with MSCT fusion as the
\emph{MSCT/MRDF detector}.

\paragraph{Multi-scale self-attention (MSSA).}
Within each MSCT block, MSSA first projects each modality into multi-head
queries, keys, and values.
The key heads are divided into four groups and processed by depth-wise
temporal convolutions with kernel sizes $1$, $3$, $5$, and $7$,
respectively.
The processed keys are concatenated before scaled dot-product attention is
computed, allowing each token to aggregate temporal evidence over multiple
receptive-field sizes.

\begin{figure}[t]
\centering
\includegraphics[width=\linewidth]{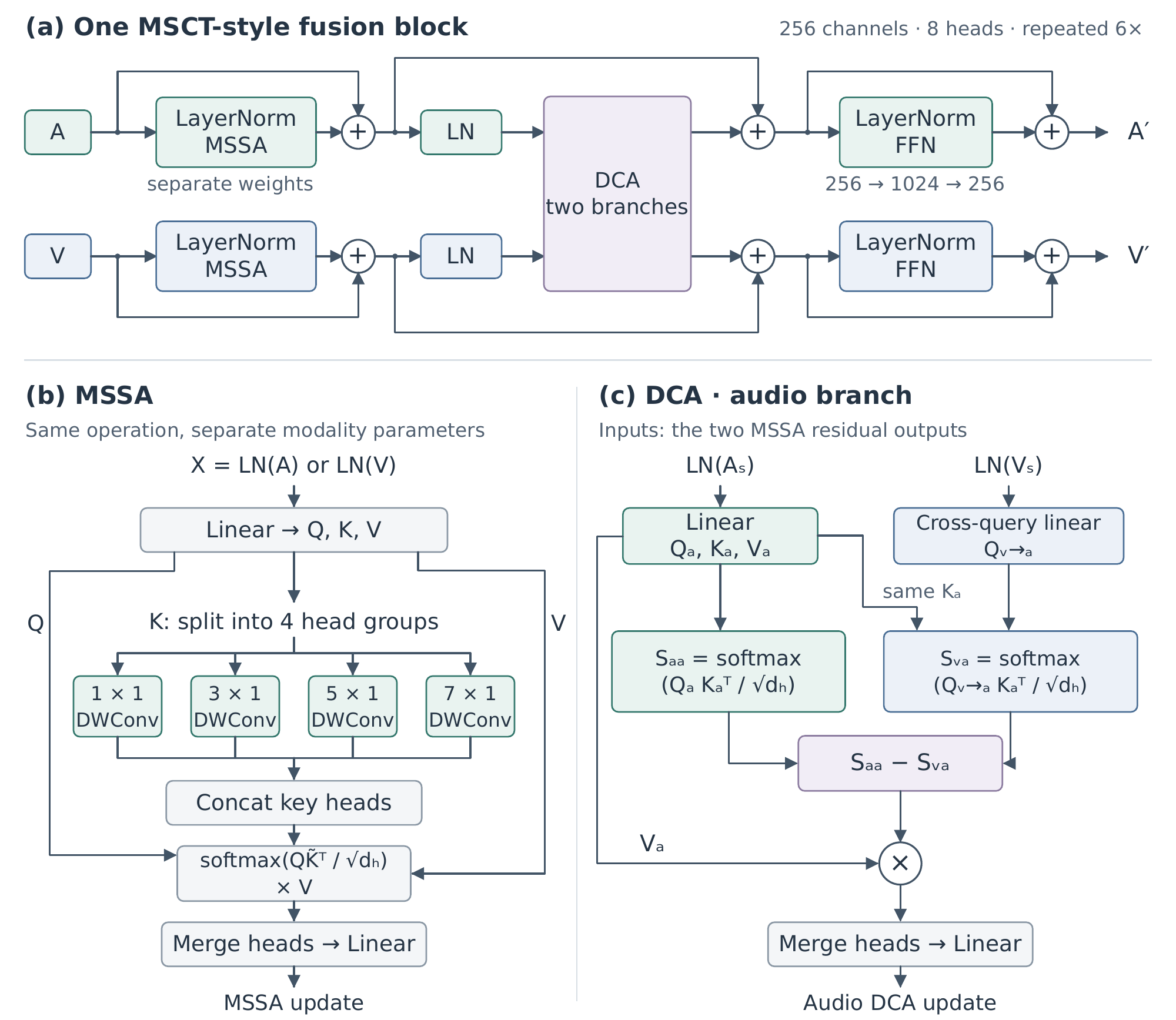}
\caption{MSCT-style fusion in this implementation, following
\cite{Wei2026MSCT}. (a) One of six fusion blocks with pre-LayerNorm
residual paths. (b) MSSA applies four temporal convolutions to key-head
groups. (c) Audio DCA subtracts cross-query attention from self-query
attention after separate softmax operations, then applies audio values;
the visual branch is symmetric. Dropout and masks are omitted.}
\label{fig:msct}
\end{figure}

\paragraph{Differential cross-modal attention (DCA).}
DCA contrasts the intra-modal attention pattern with the attention pattern
induced by the other modality.
For the audio branch, this operation can be written as
\[
\mathbf Z_A =
\left[
\operatorname{softmax}
\left(
\frac{\mathbf Q_A\mathbf K_A^\top}{\sqrt{d_h}}
\right)
-
\operatorname{softmax}
\left(
\frac{\mathbf Q_{V\rightarrow A}\mathbf K_A^\top}{\sqrt{d_h}}
\right)
\right]\mathbf V_A ,
\]
where $\mathbf Q_{V\rightarrow A}$ is a cross-query projected from the
visual stream.
The visual branch performs the symmetric operation.
DCA therefore emphasizes the discrepancy between attention supported
internally by one modality and that induced by the other modality.
Each MSCT block combines MSSA, DCA, and a modality-specific feed-forward
network through residual connections.
Figure~\ref{fig:msct} describes the implemented MSSA and DCA modules,
following MSCT~\cite{Wei2026MSCT}, and their residual connections.
Each block uses separate pre-LayerNorm residual paths for MSSA, DCA, and
the feed-forward network; both DCA branches use the same pair of
pre-normalized inputs. The first key group also uses a learned $1\times1$
convolution. There are eight heads, 256-dimensional tokens, and a
$256\!\rightarrow\!1024\!\rightarrow\!256$ feed-forward network.
The crop in Figure~\ref{fig:method}a matches audio and visual token counts;
it does not represent resampling onto a common physical-time grid.

\subsection{Synchronized-Prefix Detection}
\label{prefix}

We evaluate saved checkpoints on the first synchronized second and on the
complete available test clip. Each input contains only samples within the
requested prefix, with the original seen-validation threshold retained.
The common AV duration ranges from $1.895$ to $2.175$\,s (median
$2.015$\,s), so nominal $4$-s and $8$-s requests use the same available
content as the native-clip evaluation.

\begin{table}[h]
\centering
\caption{
First-second and native-clip detection using one checkpoint per frontend.
}
\label{tab:prefix}
\small
\setlength{\tabcolsep}{5.5pt}
\begin{tabular}{lrrrr}
\toprule
Frontend
& 1-s mAUC $\uparrow$
& 1-s Worst $\uparrow$
& Native mAUC $\uparrow$
& Native Worst $\uparrow$\\
\midrule
Spectrogram
& 80.65 & \textbf{75.07} & 84.30 & \textbf{80.12}\\
W2V2-L12
& 83.35 & 65.54 & 84.18 & 69.08\\
W2V2-L6-last
& 82.67 & 63.43 & 84.80 & 66.25\\
W2V2-L6-weighted
& \textbf{86.85} & 68.84 & \textbf{89.60} & 75.36\\
\bottomrule
\end{tabular}
\end{table}

Table~\ref{tab:prefix} shows that weighted L6 reaches $86.85\%$ mAUC
from the first second, compared with $89.60\%$ on the native clip.
Hunyuan benefits from the longer input: its AUROC increases from $68.84\%$
to $75.36\%$. Spectrogram still has the higher worst-generator AUROC at
both input durations.

\subsection{Dual-SSL Feature Fusion}
\label{dual_ssl}

Both configurations use frozen 12-layer WavLM Base+ and W2V2 Base
final-layer representations of the same waveform.
The two streams are independently projected from 768 to 256 dimensions,
aligned to the WavLM temporal grid, and normalized, yielding
$\mathbf X$ and $\mathbf Y$, respectively.
Concatenation applies a learned projection:
\begin{equation}
 \mathbf H_{\mathrm{cat}} = P_c[\mathbf X;\mathbf Y].
\end{equation}
The one-way cross-attention variant follows the residual fusion pattern
of~\cite{ElKheir2025TwoViews}, using eight heads:
\begin{equation}
 \mathbf H_{\mathrm{xattn}}
 =
 \operatorname{LN}\!\left[
   \mathbf X + \operatorname{MHA}(\mathbf X,\mathbf Y,\mathbf Y)
 \right].
\end{equation}
Here, MHA takes query, key, and value inputs in that order:
WavLM supplies the query input, and W2V2 supplies the key and value inputs.
The residual carries the prepared WavLM features $\mathbf X$.
Both fusion outputs receive a final normalization before entering the
AV detector.
\begin{table}[h]
\centering
\caption{
Single-run comparison of full-depth SSL frontends.
mAP denotes generator-macro average precision. Latency uses a separate
full-clip profiling session from Table~\ref{tab:latency_full}.
}
\label{tab:dual_ssl}
\small
\setlength{\tabcolsep}{4.0pt}
\begin{tabular}{lrrrrrr}
\toprule
Frontend
& mAUC $\uparrow$
& mAP $\uparrow$
& mEER $\downarrow$
& Worst $\uparrow$
& p50 ms $\downarrow$
& Peak MB $\downarrow$\\
\midrule
W2V2-L12
& 84.18 & 83.65 & 21.72 & 69.08 & 64.88 & 937.95\\
WavLM-L12
& 80.11 & 80.23 & 27.77 & 70.56 & 84.11 & 938.00\\
Dual concat
& \textbf{89.00} & \textbf{88.57} & \textbf{17.46}
& \textbf{76.97} & 91.24 & 1317.01\\
Dual cross-attn.
& 84.72 & 85.34 & 21.87
& 69.09 & 92.30 & 1317.51\\
\bottomrule
\end{tabular}
\end{table}

\begin{table}[h]
\centering
\caption{
Unseen-generator AUROC for the two fusion rules in the same single-run
comparison.
}
\label{tab:dual_ssl_generators}
\small
\setlength{\tabcolsep}{7.0pt}
\begin{tabular}{lrrr}
\toprule
Fusion & AniPortrait & Hunyuan & MegActor-S\\
\midrule
Concatenation
& \textbf{91.23} & \textbf{76.97} & \textbf{98.79}\\
Cross-attention
& 88.94 & 69.09 & 96.15\\
\bottomrule
\end{tabular}
\end{table}

Table~\ref{tab:dual_ssl} shows that concatenation reaches $89.00\%$ mAUC,
compared with $84.18\%$ for W2V2 and $80.11\%$ for WavLM alone.
The evaluated cross-attention configuration reaches $84.72\%$ and has
slightly higher p50 latency than concatenation ($92.30$ versus $91.24$\,ms).
Table~\ref{tab:dual_ssl_generators} shows lower cross-attention AUROC on
all three unseen generators, with the largest gap on Hunyuan ($69.09\%$
versus $76.97\%$).
The two fusion rules produce correlated scores (Spearman $.815$) but disagree
on $14.85\%$ of validation-threshold decisions on the test set. The changed
predictions therefore do not translate into better ranking than concatenation
in this comparison.

\subsection{Depth, Layer Weighting, and Generator-Wise Performance}
\label{broader}

\begin{table}[h]
\centering
\caption{
Single-run depth and weighting comparison.
p50 values come from earlier full-clip profiling rather than the
duration-dependent measurements in Table~\ref{tab:latency_full}.
}
\label{tab:broader}
\small
\setlength{\tabcolsep}{6.0pt}
\begin{tabular}{lrrr}
\toprule
Configuration
& mAUC $\uparrow$
& mEER $\downarrow$
& p50 ms $\downarrow$\\
\midrule
W2V2-L4-last & 90.57 & 14.89 & 43.87\\
W2V2-L6-last & 84.80 & 21.03 & 45.72\\
W2V2-L9-last & 81.52 & 23.61 & 48.91\\
W2V2-L12 & 84.18 & 21.72 & 64.88\\
\midrule
W2V2-L6-uniform & 84.96 & 20.33 & 45.51\\
W2V2-L6-global & \textbf{91.86} & \textbf{14.16} & 45.69\\
W2V2-L6-audio-conditioned & 90.78 & 15.24 & 46.21\\
W2V2-L6-weighted & 89.60 & 16.08 & 46.06\\
\bottomrule
\end{tabular}
\end{table}

Table~\ref{tab:broader} shows that executing more SSL layers does not
consistently improve detection: last-layer mAUC decreases from $90.57\%$
at four layers to $81.52\%$ at nine, then rises to $84.18\%$ at twelve.
At six layers, learned global weighting reaches $91.86\%$, compared with
$84.80\%$ for the last layer and $84.96\%$ for uniform averaging.
All five six-layer weighting variants have p50 latency between $45.51$ and
$46.21$\,ms. In this comparison, the choice of layer mixture changes
detection performance substantially with little change in runtime.

\begin{table}[h]
\centering
\caption{
Mean AUROC by unseen generator over three training runs.
}
\label{tab:generator_means}
\small
\setlength{\tabcolsep}{6.0pt}
\begin{tabular}{lrrrr}
\toprule
Frontend
& AniPortrait
& Hunyuan
& MegActor-S
& Worst\\
\midrule
Spectrogram
& 90.38 & \textbf{79.68} & 82.52 & \textbf{79.68}\\
W2V2-L12
& 87.21 & 76.12 & 98.23 & 76.12\\
W2V2-L6-last
& 92.28 & 74.28 & 97.67 & 74.28\\
W2V2-L6-weighted
& \textbf{94.04} & 78.58 & \textbf{99.19} & 78.58\\
\bottomrule
\end{tabular}
\end{table}

Table~\ref{tab:generator_means} shows where the mean detection gains occur.
Relative to Spectrogram, weighted L6 raises AUROC from $90.38\%$ to
$94.04\%$ on AniPortrait and from $82.52\%$ to $99.19\%$ on MegActor-S.
On Hunyuan, it obtains $78.58\%$, compared with $79.68\%$ for Spectrogram.
The macro improvement thus comes from the first two generators, while
Hunyuan remains the lowest-scoring generator for weighted L6.

\subsection{Prediction Overlap}
\label{overlap}

\begin{table}[h]
\centering
\caption{
Single-run prediction overlap using validation-selected thresholds.
Rescue is the probability that the first model is correct given that the
reference model is wrong.
}
\label{tab:overlap}
\small
\setlength{\tabcolsep}{4.5pt}
\begin{tabular}{lrrrr}
\toprule
Comparison
& Spearman
& Disagree
& Error Jaccard
& Rescue\\
\midrule
L6-weighted vs.\ L6-last
& .852 & 13.35\% & .514 & 39.24\%\\
L12 vs.\ Spectrogram
& .656 & 23.05\% & .403 & 46.02\%\\
Dual concat vs.\ L12
& .838 & 14.11\% & .532 & 36.65\%\\
Dual concat vs.\ cross-attn.
& .815 & 14.85\% & .505 & ---\\
\bottomrule
\end{tabular}
\end{table}

Table~\ref{tab:overlap} shows that weighted L6 corrects $39.24\%$ of the
examples misclassified by L6-last, and their decisions differ on $13.35\%$
of test examples.
The generator-wise breakdown gives correction rates of $58.81\%$ on
AniPortrait, $67.16\%$ on MegActor-S, and $23.76\%$ on Hunyuan.
The lower correction rate on Hunyuan is consistent with its lower AUROC in
Table~\ref{tab:generator_means}.

\end{document}